# Optical properties and carrier dynamics in Co-doped ZnO nanorods

Aswathi K. Sivan[1], Alejandro Galán-González[2,3], Lorenzo Di Mario[4], Nicolas Tappy[5], Javier Hernández-Ferrer[6], Daniele Catone[4], Stefano Turchini[4], Ana M. Benito[6], Wolfgang K. Maser[6], Simon Escobar Steinvall[5], Anna Fontcuberta i Morral[5,7], Andrew Gallant[2], Dagou A. Zeze[2,8], Del Atkinson[3], and Faustino Martelli[1]

[1]*Istituto per la Microelettronica e i Microsistemi (IMM), CNR, I-00133, Rome, Italy*

[2]*Department of Engineering, Durham University, South Rd, Durham, DH1 3LE, UK*

[3]*Department of Physics, Durham University, South Rd, Durham, DH1 3LE, UK*

[4] *Istituto di Struttura della Materia-CNR (ISM-CNR), Division of Ultrafast Processes in Materials (FLASHit), Area della Ricerca di Roma 2 Tor Vergata, Via del Fosso del Cavaliere 100, 00133 Rome, Italy.*

[5]*Laboratoire des Matériaux Semiconducteurs, Institute of Materials, Faculty of Engineering, École Polytechnique Fédérale de Lausanne, 1015 Lausanne, Switzerland*

[6]*Instituto de Carboquímica (ICB-CSIC), C/ Miguel Luesma Castán 4, 50018 Zaragoza, Spain*

[7]*Institute of Physics, Faculty of Basic Sciences, École Polytechnique Fédérale de Lausanne, Lausanne, Switzerland*

[8]*ITMO University, St. Petersburg, 197101, Russia*

## Abstract

The controlled modification of the electronic properties of ZnO nanorods via transition metal doping is reported. A series of ZnO nanorods were synthesized by chemical bath growth with varying Co content from 0 to 20 atomic % in the growth solution. Optoelectronic behavior was probed using cathodoluminescence, time-resolved luminescence, transient absorbance spectroscopy, and the incident photon-to-current conversion efficiency (IPCE). Analysis indicates the crucial role of surface defects in determining the electronic behavior. Significantly, Co-doping extends the light absorption of the nanorods into the visible region, increases the surface defects, shortens the non-radiative lifetimes, while leaving the radiative lifetime constant. Furthermore, for 1 atomic % Co-doping the IPCE of the ZnO nanorods is enhanced. These results demonstrate that doping can controllably tune the functional electronic properties of ZnO nanorods for applications.

Introduction

ZnO is a versatile wide bandgap semiconductor with a band-gap energy of about 3.3 eV at room temperature.[1–4] It has several important applications in light emitting devices, gas-sensors, bio-sensors etc., due to its semiconducting and piezoelectric properties.[5] The high binding energy of the exciton (60 meV)[3] ensures excitonic emission at room temperature, also making it an important material for photonic device applications.[6] In addition, ZnO is an earth abundant material that can be easily synthesized using several inexpensive and industrially scalable growth techniques.[7,8] The doping of ZnO with transition metals has received significant research interest in the recent years because of the addition of magnetic functionality to ZnO at room temperature.[9–11] Cobalt-doped ZnO, is doubly interesting because of both the magnetic properties and the potential tunability of the optoelectronic properties with respect to the undoped material, with the latter enabling an extension of light absorption into the visible region.[12]

The thermodynamically stable crystal phase of ZnO is wurtzite and, because of the electronegativity difference between the Zn and oxygen ions, the valence band of the ZnO is mainly governed by the p orbitals of oxygen, while the conduction band is controlled by either s or sp hybridized orbitals of Zn.[8,13] When ZnO is doped with Co, the $Co^{2+}$ competitively substitutes the $Zn^{2+}$ ions during growth and, thus the bandgap is narrowed by the sp-d exchange interaction between the band electrons of ZnO and the d-electrons of $Co^{2+}$.[12,14]

The applications drivers of cleaner, more efficient and viable energy sources have defined this research topic for over a decade.[15] In this regard hydrogen is considered to be the clean fuel of the future because the combustion of hydrogen produces large amounts of energy with water as the only by-product.[16] Currently hydrogen is mainly produced by steam reforming of methane, however, this has $CO_2$ as a by-product[17,18] and therefore cannot be considered environmentally friendly. In contrast, photoelectrochemical water splitting provides an alternative clean way to produce hydrogen using solar energy.[19–21] In a photoelectrochemical cell (PEC) used for water splitting, the efficiency of fuel production strongly depends upon the material used as the photosensitizer or the photoelectrodes. Due to its earth abundance, low cost of production, and high electron mobility, ZnO is a viable candidate as a photoanode material.[8,22] However, its wide bandgap leads to poor absorption of visible light, currently limiting the efficiency of its application as photoanode. As previously mentioned, doping of ZnO with Co has important implications in the optoelectronic properties, including the redshift of the optical absorption edge.[12] This functionality can be used to improve the efficiency of ZnO as a photoanode by increasing its absorption of the visible light.

For an efficient PEC for water splitting, a material system is needed that efficiently absorbs within the visible part of the spectrum in order to increase the light-to-current conversion in the device. Furthermore, since the absorption process is associated with the surface region a route to improve the surface to volume ratio will enhance functional device performance. A viable solution is to use a 3D array of nanowires or nanorods (NRs), because the small diameters introduce a large surface-to-volume ratio and along with the high aspect ratio offer more surface area for the interaction of light with the photoanode. In this way, it is possible to ensure higher light absorption[23,24] for the same amount of material used as compared to bulk or thin films equivalents.

In this work we report a detailed study of the optical properties of the Co-doped ZnO NRs as a function of the level of Co doping, defined here by the ionic concentration in the growth solution. In particular, the steady-state and transient optical properties of Co-doped ZnO NRs are probed through cathodoluminescence (CL), femtosecond transient absorption spectroscopy (FTAS) and time resolved photoluminescence (TRPL) measurements. Finally, photoelectrochemical measurements including wavelength dependent measurements of the incident photon to current conversion efficiency (IPCE) are reported as a function of Co concentration, showing how the doping influences the IPCE with respect to the undoped ZnO NRs and establishes the optimum $Co^{2+}$ growth concentration for PEC applications. While this paper focuses on the optical properties and carrier dynamics in Co-doped ZnO NRs, it has been demonstrated, in a complementary study that combining ZnO NRs and Co-doping in conjunction with a transparent functionalization of metal-organic framework leads to a more efficient photoelectrochemical cell.[25]

Experimental

The ZnO NRs were grown on the textured seed layers by chemical bath deposition on a 25 nm thick ZnO seed layer deposited by atomic layer deposition on a range of different substrates such as quartz, ITO-coated glass and silicon.[26] The seed layers were immersed in a solution containing equimolar concentrations (25 mM) of zinc nitrate hexahydrate and hexamethylenetetramine in water, which was subsequently heated to 90°C for 6 hours. Afterwards, the as-grown NRs were thoroughly washed with deionized water and blow dried under a $N_2$ stream. For the cobalt doping, the desired amount of cobalt nitrate hexahydrate was added to the growth solution, with the nominal doping levels being defined by the concentration ratio of the $Co^{2+}/Zn^{2+}$ in the solution, as 1%, 5% and 20%. Fig. 1 (a) presents the schematic diagram of the growth process. It is important to note that the actual levels of Co-doping will be different from the nominal values. The real density of the Co incorporated into the ZnO lattice as a dopant will be substantially lower than the nominal values.[27,28] STEM-EDS measurements (not shown) indicate indeed that the Co content is below the limit that would allow a reliable evaluation. The accurate quantification of the actual values of Co-doping is beyond the scope of this work. The effective incorporation of Co in the ZnO matrix for 1% Co

$^{2+}$ in solution has been demonstrated by XRD measurements in our previous work.[25] As it will be clear in the following, the optical properties of the samples change according to the nominal Co content increase. Hence, in line with literature,[29] we will compare the influence that different nominal values of Co$^{2+}$ in growth solution have on the optoelectronic properties of the ZnO NRs. All samples will be referred to, from now on, always by their nominal doping levels in the growth solution.

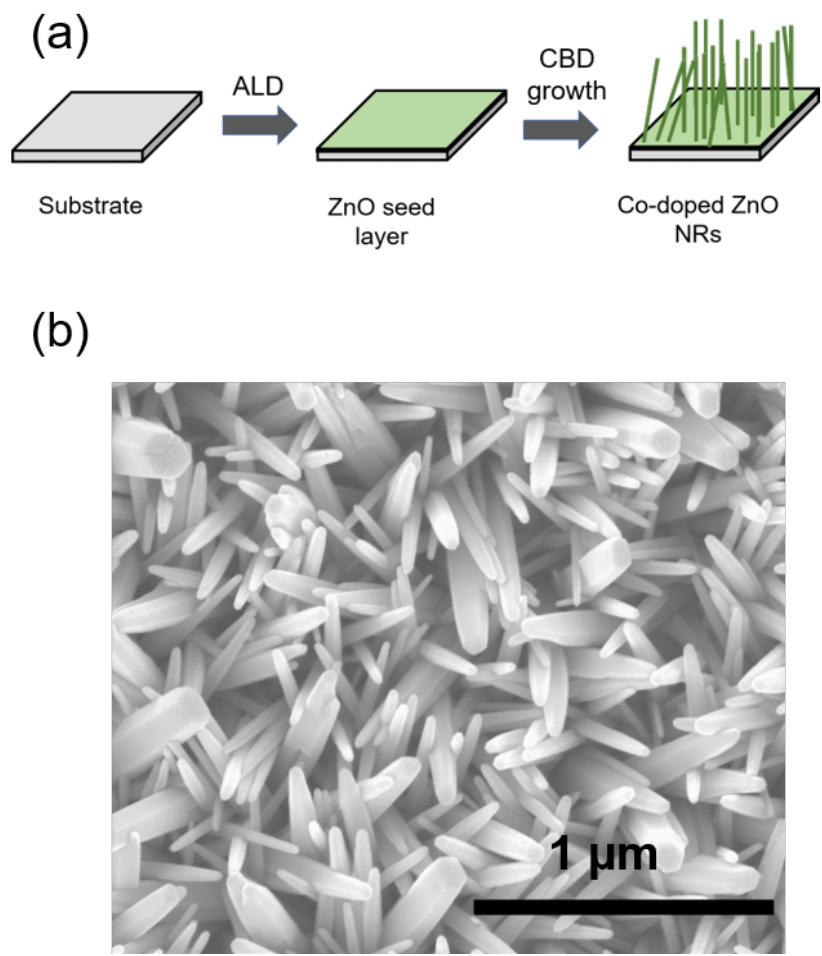

Figure 1. (a) Schematic diagram of the growth process; (b) SEM image of 1% Co-doped ZnO NRs grown by chemical bath deposition.

Fig. 1(b) shows the image of as-grown 1% Co-doped ZnO NRs on an ITO substrate taken with a scanning electron microscope (SEM). The cathodoluminescence (CL) spectra of individual NRs grown on quartz were measured by mechanically transferring them onto a conductive Si substrate to avoid charge buildup during the CL measurements using an *Attolight Allalin* integrated CL-SEM system. The samples were excited with an electron beam with energy of 5 keV at room temperature and also at 10 K. The instrument allows simultaneous SEM imaging with hyperspectral CL maps. For FTAS measurements, as-grown samples on quartz were measured in transmission at room temperature in a pump-probe configuration. The experiments were carried out with the help of a laser system consisting of Ti: Sapphire oscillator and a regenerative amplifier giving an output of 4 mJ, 35 fs wide laser pulses at 800 nm with a repetition rate of 1 kHz. The major portion of the 4 mJ, 800 nm output of the amplifier passes through an optical parametric amplifier (OPA) with a tunable output.

The OPA output was used as the pump and a white light supercontinuum, generated in a femtosecond transient absorption spectrometer of IB Photonics (FemtoFrame II), was used as the probe. The white light supercontinuum is generated for the visible region focusing 3 µJ of 800 nm in a $CaF_2$ crystal. For the samples used here, a pump of 4.51 eV (275 nm) with an excitation intensity of about 260 µJ/ $cm^2$ was used. The instrument response function was measured to be approximately 70 fs. Further details of the FTAS setup can be found elsewhere.[30] The measured quantity is the difference in absorbance (ΔA) of the probe transmitted through the excited sample and that transmitted through the unperturbed sample, as a function of the delay time between the pump and probe. The photoluminescence (PL) lifetimes were measured by time-correlated single-photon counting (TCSPC), using a photomultiplier detector and exciting the sample with a pump of 4.51 eV from the OPA at room temperature. The amperometric measurements were carried out to measure the incident photon to current (IPCE) efficiency against a reference reversible hydrogen electrode (RHE). RHE is a standard hydrogen electrode that can be used directly in an electrolyte for electrochemical measurements. A Pt electrode was used as the counter electrode and a HgO/Hg electrode as the reference. The as-grown ZnO NRs prepared on ITO/glass were used as the working electrode, i.e. the photoanode in this case. All the measurements were carried out in a pH 13 solution (NaOH in water). The potentiostat used was an Autolab PGSTAT302N, with illumination coming from a LOT ORIEL solar simulator (LS0106) equipped with a Xe arc lamp that generates an AM 1.5G illumination using a power of 300 mW cm$^{-2}$. For the IPCE measurements, a LOT Oriel MSH-300 monochromator was used.

Results and Discussion

Fig. 2 shows the CL spectra at RT of ZnO NRs. These spectra consist of a narrow UV peak centered around 3.28 eV and a broad-band spanning 1.5 to 2.7 eV. From comparison with PL measurements of Co-doped ZnO films [31–33], the narrow UV peak is assigned to the excitonic, near band-gap emission (NBE) and the broad-band luminescence to emissions related to different defects. The spectra of undoped and doped samples are similar but differ in the relative intensities of the excitonic and defect-related emissions. Whilst some debate remains regarding the origin of the defect-related emission in ZnO, [31,32] this is generally attributed to oxygen related defects. In ZnO films, previous reports have shown that with increased Co-doping, the luminescence from the excitonic recombination diminishes and the contribution from the broad band increases.[33] Fig. 2 (a) and (d) show the SEM images of NRs of 1 % Co doping and 5% Co-doping respectively. Four points are numbered along the length of each NR as 1, 2, 3, and 4. Figs. 2(c) and 2(f) show the CL spectra extracted from these points. All the 4 points taken along the NRs show two prominent peaks, the NBE peak in the UV region of the spectra and a broad defect peak in the visible region of the spectra. The ratios of intensities of NBE to that of defect band is higher for 1% Co-doped compared to 5% Co-doped NR. Figs. 2(b) and 2(e) show the panchromatic maps of the CL intensity for 1% and 5% Co-doped NRs respectively, taken under similar excitation conditions. Here the luminescence from each pixel was measured and the integrated CL intensity is depicted along the NR. From the panchromatic CL maps, it can be seen that as doping concentration increases, the intensity of the emission decreases. Also, the luminescence intensity is stronger towards the centre of the wire and weaker towards the edges. Fig. 2(g) shows the average CL emission spectra from undoped (black) and nominally 1% (red), 5% (blue), and 20% (magenta) Co-doped ZnO NRs, normalized to the NBE CL intensity. The relative contribution of the defect band with respect to the NBE peak increases with increasing Co-doping concentration in the solution.

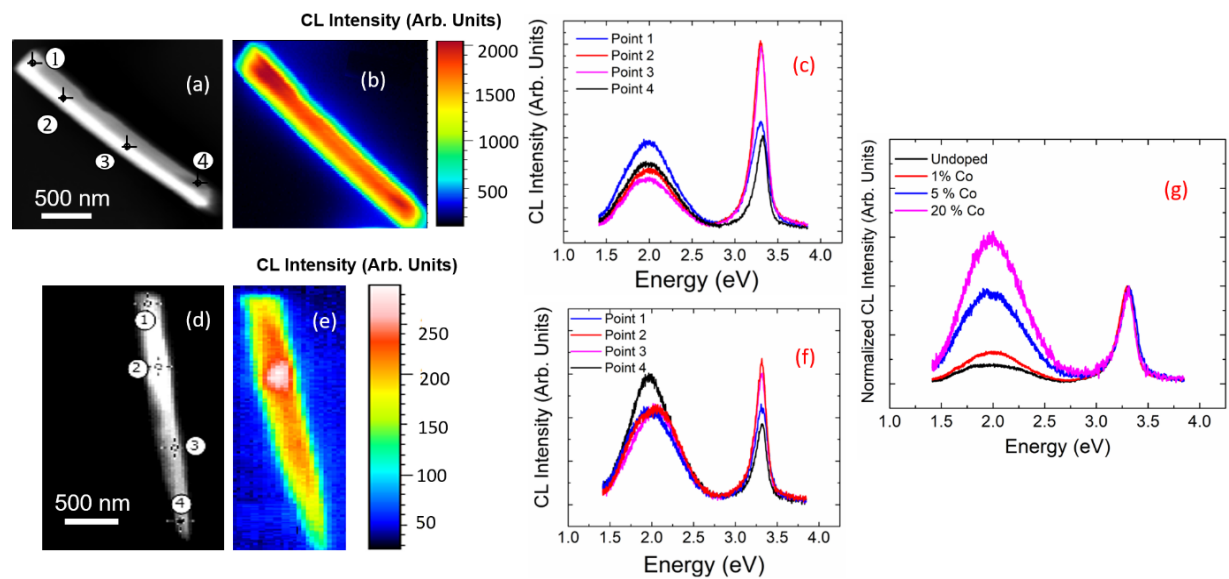

Figure 2. Cathodoluminescence at room temperature. (a) SEM image of a single 1% Co-doped ZnO NR with 4 points marked; (b) The panchromatic CL map for a 1% Co- doped ZnO NR; (c) CL spectra from the 4 points marked in (a); (d) SEM image of a single 5% Co-doped ZnO NR with 4 points marked; (e) The panchromatic CL map for a 5% Co- doped ZnO NR; (f) CL spectra from the 4 points marked in (d);(g) The CL spectra of undoped, 1%, 5% and 20% Co-doped ZnO NRs.

So far, a relative decrease in the intensity ratio between the NBE and defect luminescence is observed with increased doping. Fig. 3 illustrates in more detail, the spatial distribution of the spectral features for a 1% Co-doped ZnO NR, measured at 10 K. In particular, Fig.3(a) shows the distribution of the excitonic recombination energy, while Fig.3(b) shows that of the energy of the defect band. The UV peak shift spans 40 meV with an uncertainty (1 sigma) between 1 and 6.5 meV. The peak in the visible region of the spectra has a peak shift spanning approximately 45 meV with an uncertainty (1 sigma) below 1 meV. The peaks were fit with the help of hyperspy[34], using a split Lorentzian function for the UV peak and a Gaussian function for the visible peak. The peak energies in the UV and visible for 1% Co-doped sample are representative of the other doped samples as well.

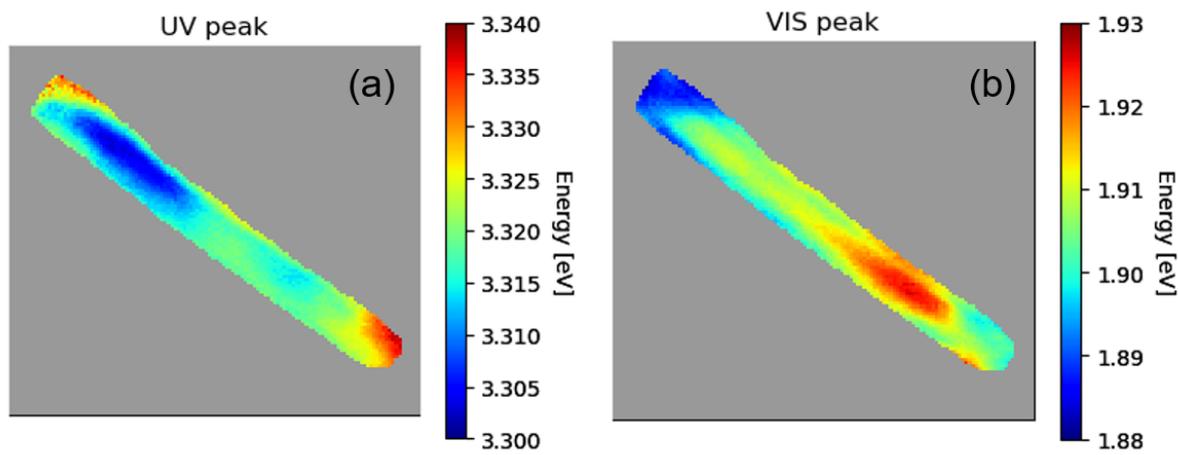

Figure 3. The hyperspectral map of CL at 10 K for a 1% Co-doped NR showing (a) the spatial distribution of the NBE peak energy; (b) same for the peak of the defects band.

The typical defects present in ZnO are vacancies, interstitials and antisites of both zinc and oxygen[35]. Oxygen vacancies can exist in 3 states, neutral ($V^0_O$), singly charged ($V^+_O$) and doubly charged ($V^{++}_O$).[31] The previously discussed broad emission centered around 2 eV is usually assigned to oxygen vacancy and Zn interstitial related transitions.[31,36–39] In the samples here, the relative increase of the broad band intensity and hence, of oxygen vacancies and Zn interstitials could be due to the substitution of $Zn^{2+}$ by $Co^{2+}$, which has a smaller ionic radii.[40] Similar CL emissions due to oxygen vacancies were observed in ZnO doped with IIIa elements as donors or with Li as an acceptor by Ohashi et al.[41] One should note here that the optical properties of high quality semiconductors are sensitive to very low densities of defects/doping.[42]

Other reports have shown that surface defects, especially oxygen vacancies, play a very important role in photocatalytic activity, especially in the efficiency of ZnO as a PEC material.[43–45] This is attributed to the enhanced absorption of visible light and increased carrier trapping associated with the increased defect states. Here CL maps show the lowest energy emission occurs at the surface, indicating that the introduction of Co in the growth of ZnO NRs may lead to a higher number of surface defects. Thus, a better photo-response is expected in those samples, which is confirmed by the IPCE measurements (see below).

In Fig. 4 we show the results of the FTAS measurements on undoped, 1% Co-doped and 20%-Co-doped ZnO NRs. The 2D false-colormaps show negative signals, indicating a reduction in absorbance, as indicated by the blue tone colors. The reduction of absorbance after photoexcitation by the pump results from carrier (de)population of the (valence) conduction band. Figs. 4 (a), (b), and (c) show the 2-D false-colormaps of the differential absorbance $\Delta A$ of undoped and nominally 1% Co-doped, and 20% Co-doped ZnO NRs, respectively. Figs. 4 (d), (e), and (f) show $\Delta A$ at different time delays for the same samples in (a), (b), and (c), respectively. In the undoped sample (Fig. 4(a)) we observe an absorbance bleaching signal centred at 3.3. eV. The bleaching signal reaches a maximum intensity between 1 ps and 2 ps (Fig. 4(d)). The small narrow peak at ~ 2.25 eV is an experimental artefact due to the second order signal of the pump laser in the monochromator and it is not a physically significant signal with regard to the interpretation of the NR behavior (this signal is observed in the other spectra in Fig.4). For undoped ZnO, the absorbance bleaching tail falls to zero at about 2.5 eV. A red shift is observed in the position of the bleaching signal in the 1% Co-doped sample, as shown in Fig. 4(b) and 4(e), and the absorption bleaching has maximum intensity at 3.18 eV. Fig. 4(e) also shows that the tail of the absorption bleaching extends over the entire probe range. For the 20% Co-doped NRs redshifts in the bleaching extend to 2.92 eV and the tail in the visible range becomes relatively more intense as the overall intensity of the bleaching peak diminishes (Figs. 4(c) and 4(f)).

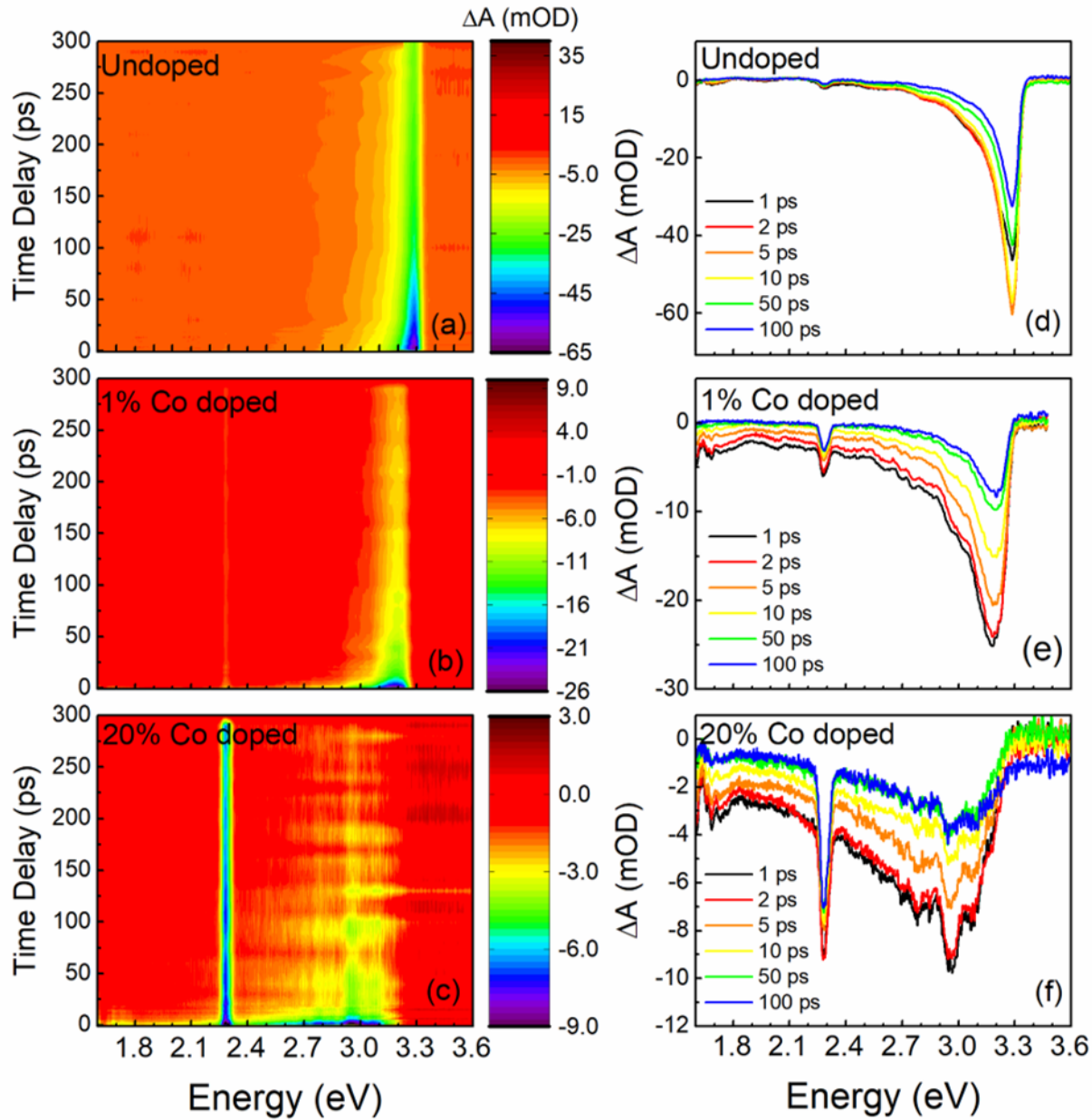

Figure 4. 2D false colormap of the difference in absorbance (ΔA) for (a) undoped, (b) 1 % Co doped and (c) 20 % Co-doped ZnO nanorods. ΔA at different time delays for (d) undoped, (e) 1 % Co-doped and (f) 20 % Co-doped NRs. All samples were excited with a pump of 4.51 eV with a fluence of 130 µJ/cm$^2$.

A further change in the absorption bleaching induced by Co-doping is a reduction of the lifetime of the NBE bleaching. Fig. 5(a) shows the temporal dependence of the absorption bleaching at different probe energies for the undoped sample. The decay of the absorption bleaching was fitted with a multiexponential curve as shown in Fig. 5(d), with three decay constants $\tau_1$, $\tau_2$, and $\tau_3$. It was observed that the shortest decay time ($\tau_1$) depends on the probe energy, see Fig. 6(b). A similar observation was made in ZnO thin films by Bauer *et al.*[39] The fit of ΔA at 3.3 eV indicates multiple decay paths with $\tau_1=20 \pm 6$ ps and $\tau_2=99 \pm 18$ ps. These two decay times relate to non-radiative recombination. The longest time constant, $\tau_3$, although included in the fitting was not determined within the precision of 300 ps wide temporal window, and has been assigned to the radiative recombination time, see below for the time-resolved PL. In the 1% Co-doped sample (see Figs. 5(b) and (d) we observe a decrease of both $\tau_1$ and $\tau_2$, being $\tau_1=5.4 \pm 0.5$ ps and $\tau_2 =82 \pm 14$ ps. For 20% Co – doped ZnO NRs we find $\tau_1=4.7 \pm 0.7$ ps and $\tau_2=32\pm14$ ps. The 5% Co-doped sample shows intermediate values for the energy shift and time constants between those doped with 1% and 20% Co.

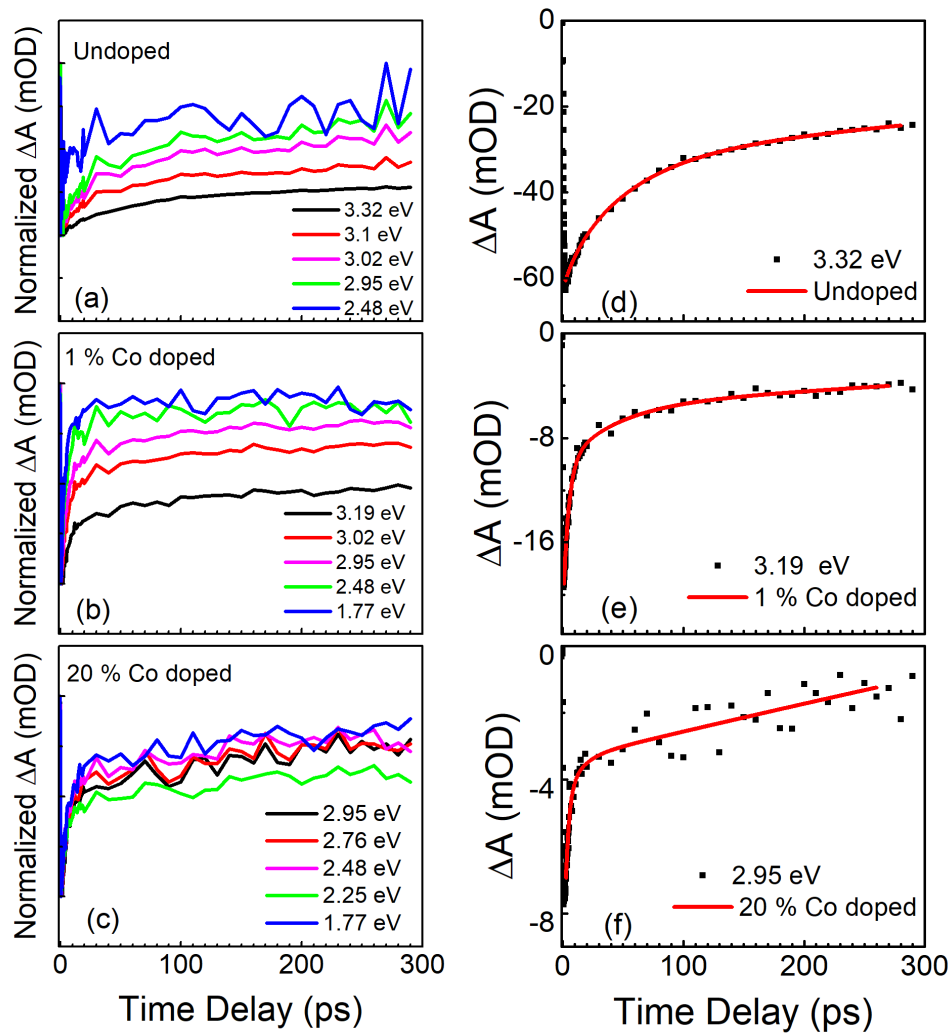

Figure 5. The time dependence of normalized ΔA at different probe energies for (a) undoped, (b) 1% Co-doped and (c) 20 % Co-doped ZnO NRs. Experimental data (black dots) and best fitting models (red line) for the NBE bleaching of (d) undoped, (e) 1% Co-doped and (f) 20% Co-doped ZnO NRs.

We attribute the red shift of the NBE absorption bleaching to an increased sp-d exchange interaction in the presence of $Co^{2+}$, leading to positive and negative corrections of the energy of valence and conduction band edges, respectively[12,14], which results in a narrowing of the bandgap. The increase in surface defects density, observed by the CL, is also reflected in the broadening of the absorption bleaching, which shows a long tail of bleaching up to 1.65 eV.

In Fig. 6 we summarize the FTAS results, with Fig. 6(a) showing the normalized ΔA at the NBE energy for all four samples (undoped, 1%-, 5%- and 20%-doped) as a function of time delay between the pump and the probe and with Fig. 6(b) illustrating the probe energy dependence of the shorter decay constant for the four samples. The reduction of the decay time constant in the doped NRs indicates increasing carrier trapping with increasing doping. Johnson *et al.* [46] have reported both transient PL and time-resolved second-harmonic generation in ZnO NRs. For high excitation intensities they observed both a sub-picosecond component and a slow component in the decay between 30 ps and 70 ps. A sub-picosecond decay component was not observed here. This fast decay component was also observed by. Sun *et al.* [47] for high excitations leading to carrier densities above the Mott density. Following these works, here the experimental condition would have led to a carrier density lower than the Mott density and hence to the formation of stable excitons.

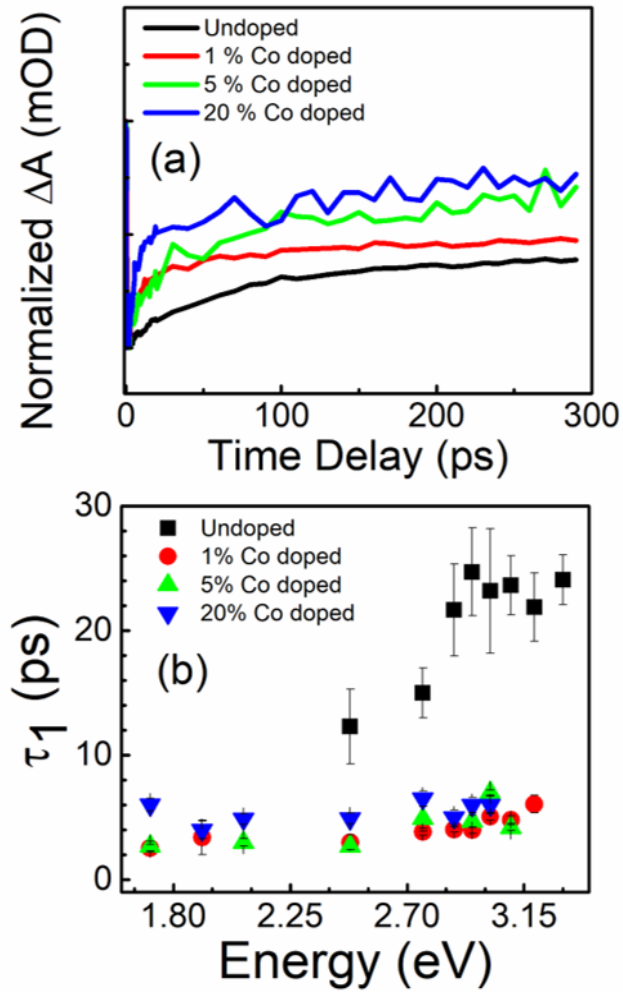

Figure 6. (a) Normalized ΔA at the bandgap bleaching energy as a function of the pump-probe delay and (b) the probe energy dependence of the decay time constant, $\tau_1$, for undoped, 1%, 5%, and 20% Co-doped ZnO NRs.

The wavelength-dependent fast decay component, $\tau_1$, which is more affected by doping than $\tau_2$, compared to undoped ZnO NRs, is attributed to carrier trapping by the oxygen vacancies, and is more efficient in the doped samples. $\tau_2$ is attributed to other non-radiative recombination channels that depend on the details of the Co-doping, such as the formation of Co clusters[48].

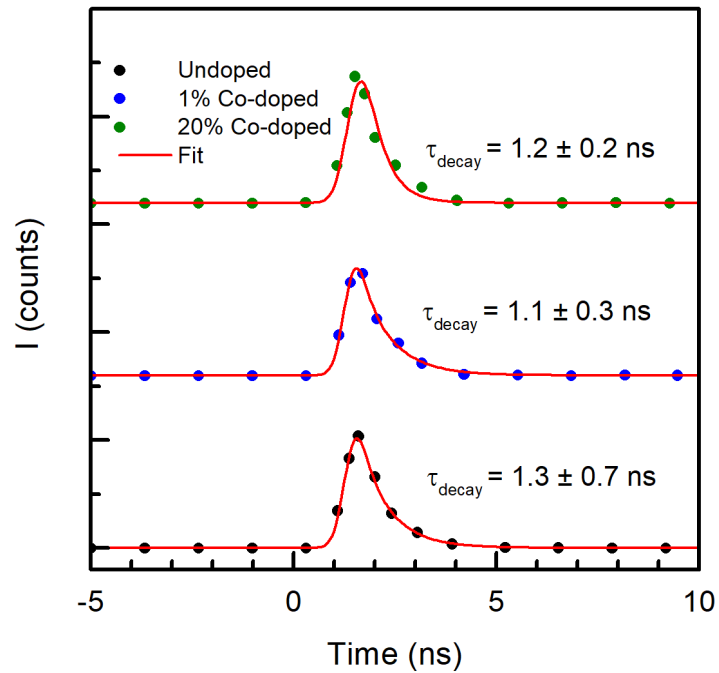

Figure 7. Time resolved PL (TRPL) measurements of undoped (black), 1% Co-doped (blue) and 20% Co-doped (green) ZnO NRs.

If the Co doping has a measurable effect on the fast carrier dynamics, its effects on the radiative lifetime are negligible. Fig. 7 shows the temporal behavior of the luminescence intensity of the NBE peak, along with the best fitting model to the data. The *x*-axis represents the time in nanoseconds and the *y*-axis represent the intensity count (I) obtained through TCSPC. The spectra obtained in PL, not reported, are very similar to those shown earlier for the CL measurements. The PL decay of the NBE emission was fitted with a double exponential , in which the best fit requires a fast component that was shorter than the response time of the instrument (0.5 ns) used for measuring time-resolved PL (TRPL), and a slow component with a decay time of $1.3 \pm 0.7$ ns for the undoped ZnO NRs. Similar values were observed by Jung *et.al*[49] in ZnO thin films from TRPL measurements. With increased doping concentration the ZnO NRs with Co, the variations of the slow component are unchanged within the experimental error, being $1.1 \pm 0.3$ ns for 1% Co doping and $1.2 \pm 0.2$ ns for 20% Co doped NRs, indicating no dependence on doping. A time constant of ~1 ns is compatible with typical radiative recombination times. However, as both the PL and CL intensities decrease with increasing doping, and the decay times in the FTAS measurements depend on doping, the constant radiative recombination time measured is relative to the residual electron-hole pair density, decreasing as the doping increases, once the carrier traps are saturated.

The CL measurements suggest an increase in the oxygen vacancy defects with the concentration of Co-doping that is related to the smaller ionic radii of $Co^{2+}$ compared to Zn. The time dependence of the absorption can be fitted well using a multiexponential model, which shows that the fast decay components become faster with increasing doping. In contrast, the radiative recombination time remained constant within the experimental errors, as expected. Such faster decay with increased doping, indicates more efficient carrier trapping that is induced by the increased surface defects in the doped NRs. In samples with high concentration of oxygen vacancies, the defect states are more delocalized and can overlap with the valence band edge, which also contributes to narrowing of the bandgap.[50] Given the high surface-to-volume ratio of NRs, the surface plays a more important role than in thin films or bulk. This is clear in the photoresponse of the doped NRs, which increases in the doped samples, as shown below, as the two excess electrons associated with each oxygen vacancy can also help improve the photocatalytic activity. Additionally, while the transient absorbance NBE signal is red-shifted, the NBE luminescence, both in CL and PL, is not. The sp-d exchange interaction responsible for the red shift is accompanied by the creation of non-radiative recombination channels, which are probably of surface origin.

Finally, the IPCE measurements show the efficiency of a photoelectrode to convert an incident photon into an electron, which is a critical quantity in the assessment the performance of a photovoltaic/photochemical material. Co-doping of ZnO NRs has a beneficial effect on the photoresponse of the material. Fig. 8(a) shows the wavelength dependent IPCE measurements at 1.33 V against the standard RHE for the undoped and Co-doped ZnO NRs, showing that the IPCE increases with the doping.

The IPCE values for all of the doped samples were higher than that obtained for the undoped sample (<40%). In particular, the highest IPCE was for 1% C-doped NRs, which was ~51 % higher at 350 nm, followed by the other doped samples that showed improvements of about 40%. Furthermore, the presence of the Co-doping greatly enhances the IPCE in the near UV region, between 300 nm and 350 nm. In addition, a small red-shift in the IPCE is observed for with doping, which corresponds with the previously detailed optical characterization. Though there is more light absorption in the visible region, this does not translate to a higher current generation in the visible part of the spectrum. This could be due to the fact that there may be an energy barrier or overpotential to overcome, and the absorption at the higher wavelengths is not sufficient to create carriers to overcome this. Fig. 8(b) shows the electronic response to light exposure of the undoped and Co-doped NRs under AM 1.5G illumination at a voltage sweep between -0.5 V and 1.33 V vs RHE. The photo-generated current density is about 40 % higher for the 1 % Co-doped samples as compared to the undoped ZnO NRs. With increasing doping, the photo-generated current density decreases (5% Co) until it is smaller than that of the undoped NRs in case of the 20% Co-doped NRs as seen from Fig 8 (b). While surface defects resulting from doping act as hole trapping sites, providing excess electrons, and improve the absorption of visible light, very heavy doping may lead to a decrease in the crystalline quality and create bulk defects, such as Co clusters[44], and increase the non-radiative recombination sites, which in turn results in a decrease in the photocurrent density.

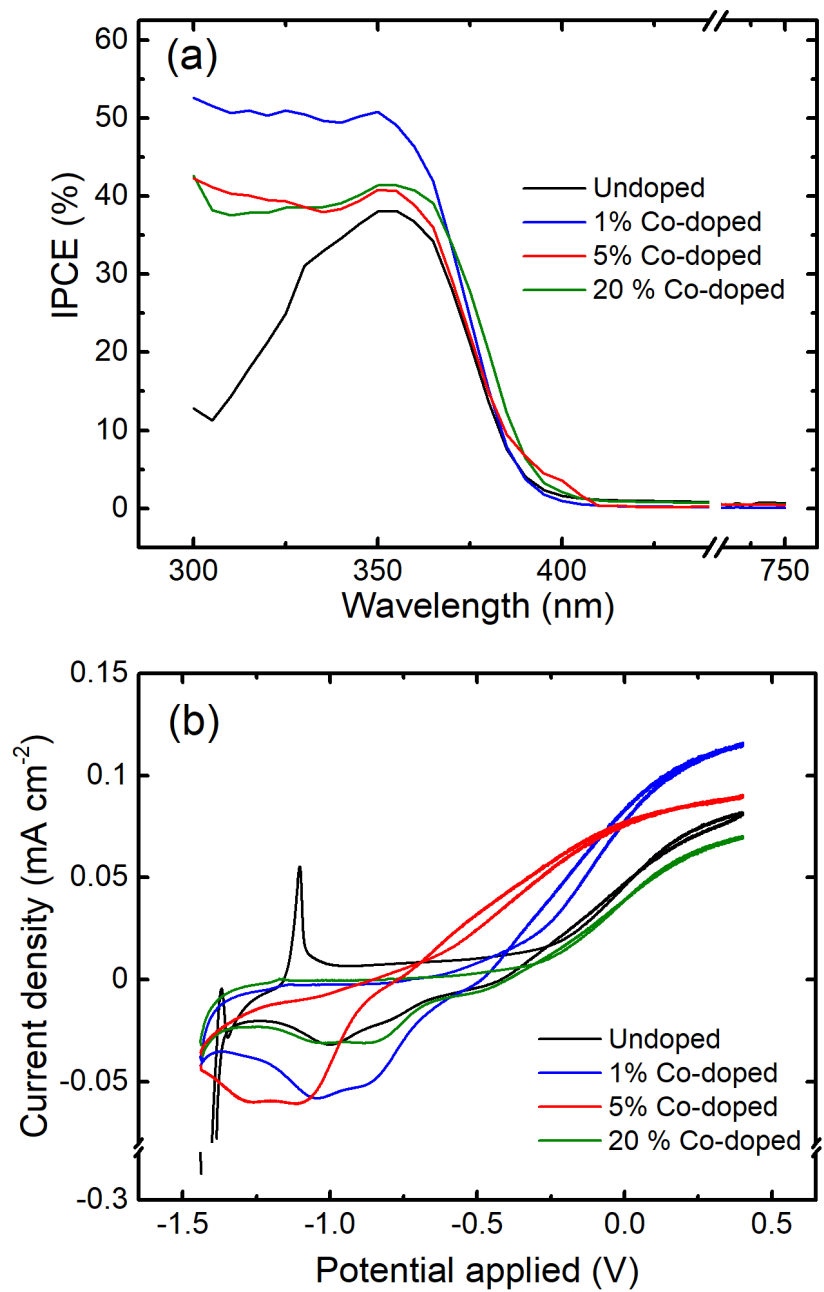

Figure 8. (a) Wavelength dependent IPCE measurements at 1.33 V vs RHE for undoped and Co-doped ZnO NRs; (b) Cyclic sweeping voltammetry under AM 1.5 G illumination. Black curves: undoped ZnO NRs; blue: 1%, red: 5%, and green: 20%- Co-doped ZnO NRs.

Conclusion

The optical properties of Co-doped ZnO nanorods have been systematically studied in order to understanding the influence of doping on the electronic properties and tested their potential for use as photoanodes. It was deduced that more oxygen vacancies are created with increased Co-doping and the luminescence intensity decreases with increasing doping due to the formation of more favorable non-radiative recombination centers. Moreover, it was shown that higher Co-doping leads to the faster decay of the bleaching signal. This could be due to efficient carrier trapping in the presence of more surface defects. On the other hand, the radiative recombination lifetime does not change with doping, although a reduced density of electron-hole pairs recombines radiatively. Co-doping also causes a red shift of the excitonic absorption bleaching. This is attributed to the sp-d exchange interaction between the d-electrons of Co and the band electrons of ZnO as well as the delocalization and overlap of the oxygen vacancy energy levels and the valence band of the ZnO. Finally, measurements show higher photo generated current density for 1% and 5% Co growth concentration values than the undoped ZnO NRs, while 20 % Co has a lower photocurrent density, due to increased rates of non-radiative recombination. All the measurements indicate the importance of the surface defects in determining the optical properties of the ZnO NRs and that the Co doping, at least at growth concentration values around 1%, affects the surface states in a beneficial way for the photo-response properties of the NRs. Doping ZnO with Co hence significantly modifies its opto-electronic properties and has important implications on the photo response of ZnO by increasing light absorption in the visible region and decreasing the carrier lifetime. The nominal 1% Co-doped ZnO NRs have shown the highest IPCE and photo-response. The findings offer a possibility to functionalize ZnO to overcome its shortcomings in its usability as a photoanode.

## Acknowledgements

This work has received funding from the Horizon 2020 program of the European Union for research and innovation, under grant agreement no. 722176 (INDEED) and SNSF via project nr-40B2-0_176680 and BSCGI0-157705. AKS acknowledges the *Laboratoire des Matériaux Semiconducteurs, EPFL* for the kind hospitality during her secondment within the INDEED project. We thank Vincenzo Grillo (CNR-Nano, Modena, Italy) for very helpful discussions. WKM, AMB acknowledge *Spanish MINEICO* under project ENE 2016-79282-C5-1-R